\documentclass[preprint,12pt]{elsarticle}
\usepackage{amssymb}

\journal{Physics Letters B}

\begin{document}

\begin{frontmatter}

\title{Relativistic Two-Body Coulomb-Breit Hamiltonian in an  External Weak Gravitational Field}

\author{J. A. Caicedo and L. F. Urrutia}

\address{Instituto de Ciencias Nucleares, Universidad Nacional Aut{\'o}noma de M{\'e}%
xico, \\
A. Postal 70-543, 04510 M{\'e}xico D.F., M{\'e}xico}

\begin{abstract}
A construction of the Coulomb-Breit Hamiltonian for a pair of fermions, considered as a quantum two-body system, immersed in an arbitrary  background
gravitational field  described by  Einstein's  General Relativity is presented.  Working with
Fermi normal coordinates for a freely falling observer in a spacetime region where there are no background
sources and  ignoring the  gravitational back-reaction of the system, the
effective Coulomb-Breit  Hamiltonian is  obtained
starting from the S-matrix element corresponding to the one-photon exchange between the charged fermionic currents. The  contributions due to retardation are considered
up to order $(v/c)^2$ and  they are subsequently written as effective operators in the relativistic quantum mechanical Hilbert space of the system.
The final Hamiltonian  includes  effects linear in  the curvature and up to order $(v/c)^2$.
\end{abstract}


\end{frontmatter}

\section{Introduction}
The connection between Einstein's General Relativity (GR), the dynamical theory
of space-time, and Quantum Mechanics (QM) has a long history in physics and more
than ever is nowadays a subject of intense research, both from  theoretical
and  experimental perspectives, mainly in connection with possible deviations from GR and/or QM. Until
now, experiments have already confirmed that inertia and Newtonian gravity affect
quantum particles, mainly electrons and neutrons, in ways that are fully
consistent with GR down to distances of the order of
${10^{-8}}\, \rm{cm}$.
Gravitational-inertial fields leave their mark on particle wave functions in a
variety of ways; particularly,  they induce quantum phases that have been measured
in some of the most renowned experiments on this topic
\cite{COW,Bonse-Wroblewski,Kasevich-Chu,Shimizuetal,Rhieleetal,
Hasselbach-Nicklaus}. Recently, the purpose of tests of gravity has been mainly focused in determining
the validity of the equivalence principle at the level of atoms, molecules, quasi-particles and antimatter.
This translates into a need  for higher precision tests of GR, and atomic physics together with quantum optics
offer some of the most accurate results and promising scenarios.

Most of the  experimental tests of GR within the quantum
realm arise  from predictions on  fully covariant wave equations where inertia and gravity appear
as external classical fields, providing very valuable information on how
Einstein's views carry through into the quantum world. However, many of the initially
proposed and tested effects considered mainly one-particle systems. This approach leaves aside the
possibility of using the internal quantum structure of atoms or molecules as additional parameters.
Nevertheless, this situation is drastically changing recently
due to  the fast development of  matter wave interferometry \cite{LNP,Dimopoulosetal,AtmInt}.

Broadly speaking, previous work related to the description of atoms in a background gravitational field can be divided in three
main branches, though with much interrelation among them:
(i) the study of gravitational modifications to atomic
spectra of mainly hydrogen-like atoms, which were considered as a quantum reduced mass revolving
around some force center
\cite{Parker1,Parker2,Parker3,Gilletal,Pint93,Audretsch,Parker97,Obukhov01,Silenko05,Zhen07a,Zhen07b,Pavlichencov09}. A two-body description of the atom can be found in Refs. \cite{Fischbach79,Fischbach-Freeman,Haugan3}.
(ii) Restricted proofs of the equivalence principle for test bodies made of classical electromagnetically interacting particles, within  the PPN  framework or within the
$TH\epsilon\mu$ formalism for static  spherically symmetric spacetimes
\cite{LL,Audretsch-Schafer,Haugan1,Haugan2}. The formulation of a quantum equivalence principle has been studied in Refs. \cite{Claus1,Claus2}.
(iii) Tests and studies of the gravitational red shift, where atomic clocks were described as quantum mechanical systems within the $TH\epsilon\mu$ or PPN formalisms \cite{Will74,Vessot80,Turneaure83,Krisher93,Krisher96,Claus98,Bauch02}.

In this letter we consider the effect of a  classical background  gravitational field,
described by GR,  upon matter at the atomic level, where the use of QM becomes
mandatory.
We extend previous work in the following aspects:
(i) We will consider the case of hydrogen-like atoms described as
two-body systems in the way first discussed in Ref. \cite{Fischbach-Freeman}. The inclusion of the position of the atom as a  new degree of freedom may provide additional
couplings to probe the effects of gravity upon quantum objects in the region where tidal forces become important. (ii) The
background gravitational field has no particular spatial  symmetry.
(iii) Following the steps of Ref. \cite{Greiner}, a quantum mechanical relativistic two-body
effective potential in the background gravitational field is produced, which takes into
account  retardation effects arising from the one-photon exchange between the
currents. This constitutes the gravitational generalization of the
relativistic Coulomb-Breit potential \cite{BREIT}.

Most of the general assumptions underlying this work are similar to those of
Refs.\cite{Parker1,Gilletal,Audretsch,Fischbach-Freeman,LL,Audretsch-Schafer}:
(i) The external gravitational field is well described by GR, so that it
satisfies Einstein's field equations. Additionally, we ignore back-reaction
effects.
(ii) All the interacting particles are influenced by the background
gravitational field at the quantum level.
(iii) There exists an ideal  freely falling observer  in a region outside the
sources of the background gravitational field. This observer determines that, in
his coordinate patch, the metric of the spacetime has the form
\footnote{Our metric has signature -2, $\eta^{{\bar \alpha}{\bar
\beta}}=diag(-1,+1,+1,+1)$,
$e_{\bar \alpha}{}^\mu{}$ is the tetrad such that
$e_{\bar \alpha}{}^\mu{}\,\eta^{{\bar \alpha}{\bar \beta}}\, e_{\bar
\beta}{}^\nu{}=g^{\mu\nu}$, ${\bar \alpha}=\left\{{\bar 0},{\bar 1},{\bar
2},{\bar 3}\right\}=\left\{{\bar 0}, {\bar a} \right\}$
are tetrad indices, while $\mu=\left\{0,1,2,3\right\}= \left\{0, a \right\}$ are
coordinate indices. Also
$e_{\bar \alpha}{}^\mu{} \, e^{\bar \alpha}{}_\nu= \delta^\mu_\nu$. The symbol
$\circeq$ denotes an equality including  at most  terms linear in the curvature.
}
\begin{equation}\label{Fermetr}
g_{\alpha\beta}=\eta_{\alpha\beta}+{ \cal Q}_{\alpha b\beta c}x^{b}x^{c}\equiv
\eta_{\alpha\beta}+h_{\alpha\beta}\left(x^{0}\right),
\end{equation}
where
\begin{equation}
{\cal Q}_{0a0b}=-\breve{R}_{0a0b}\left(x^{0}\right),\,\,
{\cal Q}_{0abc}=-\frac{2}{3}\breve{R}_{0abc}\left(x^{0}\right), \,\,
{\cal Q}_{abcd}=-\frac{1}{3}\breve{R}_{abcd}\left(x^{0}\right).
\end{equation}
Here $x^{a}$ are three spatial coordinates, $x^{0}$ is the proper time of the
observer and $\breve{R}_{\alpha\beta\gamma\delta}$ is the projection of
the background Riemann tensor on the orthonormal tetrad carried by him, which in
general depends on the proper time due  to the motion of the observer.
The weakness of the gravitational interaction induces
very small corrections upon the observables, so that in our final results we
preserve only quantities up to first order in
$\breve{R}_{\alpha\beta\gamma\delta}$.
The metric (\ref{Fermetr}) is that of a
freely falling observer using Fermi normal coordinates, which are appropriate
to describe local experiments performed by inertial observers
\cite{Manasse-Misner}.
(iv) The spatial extension and the time duration of events related to
observations in the quantum system are very small compared with the characteristic
lengths and times of appreciable changes in the observer's Riemann tensor
\cite{Audretsch}. This allows well defined energy levels and the use of
time-independent perturbation theory. Within this adiabatic approximation it is
possible to ignore  the $x^{0}$ dependence of the metric and of all the objects
constructed from it: the time coordinate becomes just a fixed but arbitrary
parameter \cite{Audretsch-Schafer}.
(v) Conditions (i) and (iii) imply that the observer determines that his Ricci tensor ${{\breve R}^\alpha{}_\beta}$ is equal to zero.
We  also assume that during the measurements performed by the observer there are no particle creation effects due to the
gravitational field.

The coupled equations that describe the electromagnetic interaction between spin
$1/2$ fermions in a background gravitational field, outside the gravitational
sources,  are
\begin{equation}\label{ecDiracQED}
 ic\hbar\gamma^{\mu}D_{\mu}\psi-mc^{2}\psi=q\gamma^{\mu}\psi
A_{\mu}, \qquad
D^{\mu}D_{\mu}A^{\nu}=q\bar{\psi}\gamma^{\nu}\psi\equiv
{J}^{\nu},
\end{equation}
where we have chosen the Lorentz gauge $D_\mu A^{\mu}=0$. Here $
D_{\mu}= \partial_{\mu}-\frac{i}{2}\omega^{BD}{}_{\mu}J_{BD}
$
is the covariant derivative written in general terms,
$\omega^{\bar{\alpha}}{}_{{\bar{\beta}}\mu}=e^{\bar{\alpha}}{}_{\delta}\left(D_{\mu}e_{\bar{\beta}}{}^{
\delta}\right)$ is the spin connection,  and $J_{BD}$ are the Lorentz group generators in the corresponding
representation for each field.

\section{The photon propagator}\label{Sec1}
The fundamental quantity required to construct an
effective relativistic Hamiltonian describing the electromagnetic interaction
between two charges is the curved space  Feynman Green function $G^{\mu}{}_{\nu'}(x',x)$ for the photon
\cite{Ohta-Kimura}.  We consider its Hadamard representation because it is valid in all spacetimes \cite{Brown-Ottewill}.
The differential equation for the electromagnetic Feynman Green function is
\begin{equation}\label{GFEMwavecu}
\square \,
G^{\alpha}{}_{\beta}(x,x')=-\kappa^{\alpha}{}_{\beta}\left(x,x'\right)
\delta_{4}\left(x,x'\right),
\end{equation}
where $\square\equiv g^{\mu\nu}\nabla_{\mu}\nabla_{\nu}$,
$\, \kappa^{\alpha}{}_{\beta}$ is the parallel propagator bitensor between $x'$
and $x$, and $\delta_{4}\left(x,x'\right)$ is the invariant Dirac delta
distribution for our observer. The solution to (\ref{GFEMwavecu}), that vanishes at infinity for arbitrary time, is
\begin{equation}\label{EMGFfinvac}
G_{\alpha\beta}(x, x')=\frac{1}{\left(2\pi \right)^{2}}\left(\frac{\kappa_{\alpha\beta}(x,x')}{\sigma(x,x')+i\epsilon}
\right).
\end{equation}
The expressions for the parallel propagator $\kappa^{\alpha}{}_{\beta}\left(x,x'\right)\equiv \, e_{{\bar \gamma}}{}^{\alpha}\left(x\right)e^{{\bar \gamma}}{}_{\beta}\left(x'\right)$ are \footnote{The  orthonormal tetrad corresponding to the metric (\ref{Fermetr})
is: $e_{\bar{0}0}=-1-\frac{1}{2}h_{00},\; e_{\bar{0}a}=-\frac{1}{4}h_{0a},\; e_{\bar{a}0}=\frac{3}{4}h_{0a},\; e_{\bar{a}b}=\eta_{ab}+\frac{1}{2}h_{ab}$ \cite{Audretsch}.}
\begin{eqnarray}\label{ProparlCNF}
\kappa_{00}& \circeq   \eta_{00}+\frac{1}{2}\left[h_{00}\left(x\right)+h_{00}\left(x'\right)\right], \qquad
\kappa_{0a}\circeq  \frac{1}{4}\left[3h_{0a}\left(x\right)+h_{0a}\left(x'\right)\right], \\
\kappa_{a0}& \circeq   \frac{1}{4}\left[h_{0a}\left(x\right)+3h_{0a}\left(x'\right)\right], \qquad
\kappa_{ab}\circeq   \eta_{ab}+\frac{1}{2}\left[h_{ab}\left(x\right)+h_{ab}\left(x'\right)\right].
\end{eqnarray}
Here $\sigma(x,x')$ is the Synge world function, given by
\begin{equation}\label{SynCNF}
\sigma\left(x,x'\right)\circeq   \frac{1}{2}\eta_{\mu\nu}\left(x-x'\right)^{\mu}\left(x-x'\right)^{\nu}
+\frac{1}{2}{\cal Q}_{\mu a \nu
b}{ \cal A}^{ab}\left(x-x'\right)^{\mu}\left(x-x'\right)^{\nu}.
\end{equation}
with
\begin{equation}\label{Aabdef}
{\cal A}^{ab}\equiv\left[\frac{1}{2}\left(x'^{a}x^{b}+x^{a}x'^{b}\right)
+\frac{1}{3}\left(x-x'\right)^{a}\left(x-x'\right)^{b}\right].
\end{equation}
Note that expression (\ref{EMGFfinvac}) contains terms of higher order than first in the
curvature, arising from the denominator $\sigma+i\epsilon$. We will keep the
complete singular structure of the Green function because, in the following calculations, we will require expressions for the corresponding poles
up to first order in  $h_{\mu\nu}$.  Let us remark that $\sigma(x,x')=\sigma({\mathbf{x}, \mathbf{x'}, x^0-x'^0})$.

\section{The one-photon  interaction}\label{Sec2}
In order to incorporate  gravitational corrections into  the electromagnetic
interaction between  two charged particles, we will use the S-Matrix method
\cite{Ohta-Kimura} as described in Ref.\cite{Greiner}, but generalized to
a slightly curved space \cite{CUPdelC}.
We need to evaluate the S-matrix element corresponding to
the exchange of one photon between the two fermionic currents located at $x_2$ and $x_1$, shown in Fig. (1),
which  is given by
\begin{equation}\label{Sfiele}
S_{fi}^{\ \left(1\right)}=\frac{1}{c\hbar}\int
J^{\mu\left(2\right)}_{fi}\left(x_2\right) G_{\mu\nu}\left(x_2,x_1\right)
J^{\nu\left(1\right)}_{fi}\left(x_1\right)\, d^{4}V_2 \, d^{4}V_1,
\end{equation}
where $G_{\mu\nu}\left(x_2,x_1\right)$ is the electromagnetic Feynman Green
function (\ref{EMGFfinvac}) and $d^{4}V\equiv\sqrt{-g\left(x\right)}\ d^{4}x$ is the invariant volume element at point $x$.
The transition
current for each charge $q_n$ has the form given in Eq. (\ref{ecDiracQED}).
As a consequence of the adiabatic approximation, the Dirac fermions which describe the charged
particles can be
represented by spinors with a definite positive frequency, in
the coordinate patch of our observer. We can write the spinor field as
\begin{equation}\label{Espinposit}
\psi\left(y\right)=e^{-ik^{0}y^{0}}\Psi \left({\mathbf y}\right) \quad {\rm with} \quad  k^{0}>0,
\end{equation}
which satisfies the generalized Dirac equation given in (\ref{ecDiracQED}). The explicit form for each  current is
\footnote{Our conventions for the gamma matrices are: ${ \gamma}^{{\bar
\alpha}}$ denote  flat-space matrices, such that $ \{ { \gamma}^{{\bar \alpha}},
{ \gamma}^{{\bar \beta}} \}= 2 \eta^{{\bar \alpha} {\bar \beta}}, \,\,
({\gamma}^{{\bar 0}})^{T*}=- { \gamma}^{{\bar 0}}, \,\, ({ \gamma}^{{\bar
\alpha}})^{T*}= { \gamma}^{{\bar \alpha}}$. We have
$\gamma^\mu(x)=e_{\bar \alpha}{}^\mu{}(x){ \gamma}^{\bar \alpha}$.}
\begin{equation}\label{corrTrans}
J^{\mu (n)}_{fi} (y)=\left[
q_{n}\left( \Psi_{f}^{(n)}(\mathbf{y})\right)^{T *}
\gamma^{\bar{0}}_{\left(n\right)}\gamma^{\mu}_{\left(n\right)}\left(\mathbf{y}
\right)\Psi_{i}^{\left(n\right)}({\mathbf
y})\right]
e^{i\Delta_{fi}^{\left(n\right)}y^{0}},
\end{equation}
where we have defined
$\Delta_{fi}^{\left(n\right)}\equiv k^{0}{}_{f}^{\left(n\right)}-k^{0}{}_{i}^{\left(n\right)}$.
Let us construct the two-particle states $\Phi_{i,f}\left(\mathbf{1},\mathbf{2}\right)$  as the
usual direct product of the spinorial wave function $\Psi^{\left(n\right)}({\mathbf
x})$ of each particle. We can then rewrite
$S_{fi}^{\ \left(1\right)}$ as a matrix element of an operator
${\cal W}^{\left(\mathbf{1},\mathbf{2}\right)}$ between the initial and final
two-particle states : $\Phi_{i}\left(\mathbf{2},\mathbf{1}\right)=\Psi^{(2)}_{i}(\mathbf{x}_2)\otimes \Psi^{(2)}_{1}(\mathbf{x}_1)$ and
$\Phi_{f}\left(\mathbf{2},\mathbf{1}\right)=\Psi^{(2)}_f(\mathbf{x}_2)\otimes \Psi^{(1)}_f(\mathbf{x}_1)$, respectively, in such a way that
\begin{equation}\label{SfiW}
S_{fi}=\int \left(\Phi_{f}\right)^{T \ast}
\left(\gamma_{(2)}^{\bar 0}\gamma_{(2)}^0\right)
\left(\gamma_{(1)}^{\bar 0}\gamma_{(1)}^0 \right)
{\cal W}^{\left(\mathbf{1},\mathbf{2}\right)}\,
\Phi_{i}\, d^4V_2 \, d^4V_1 \equiv  i \delta(\Omega)\, \langle\Phi_{f}\vert {\hat{\cal W}}_e \vert
\Phi_{i}\rangle,
\end{equation}
where $\Omega \equiv \Delta_{fi}^{\left(2\right)}+\Delta_{fi}^{\left(1\right)}$ {\bf and $\delta(\Omega)$} leads to energy conservation.
Substituting (\ref{corrTrans}) in (\ref{Sfiele}) and with help of (\ref{SfiW}) we identify the operator
\begin{equation}\label{def1potef}
{\cal W}^{\left(\mathbf{2},\mathbf{1}\right)}=\alpha_e \, \frac{\gamma^{0}_{\left(2\right)} \gamma^{\bar{\mu}}_{\left(2\right)}}
{g^{00}\left({\mathbf
x_2}\right)}\, \, {\cal G}_{{\bar \mu} {\bar \nu}}\left(\mathbf{2},\mathbf{1}\right)\frac{
\gamma^{0}_{\left(1\right)}\gamma^{\bar{\nu}}_{\left(1\right)}}{g^{00}\left({
\mathbf x_1}\right)}, \qquad \alpha_e= -\left(\frac{q_{1}q_{2}}{\hbar
c}\right),
\end{equation}
where
\begin{equation}\label{def3pe}
{\cal G}_{{\bar \mu}{\bar \nu}}({\mathbf x_2},{\mathbf
x_1})= \frac{\eta_{{\bar \mu}{\bar \nu}}}{\left(2\pi\right)^{2}}\int\int\frac{e^{
i\left(\Delta_{fi}^{\left(2\right)}
x_2^{0}+\Delta_{fi}^{\left(1\right)}x_1^{0}\right)}}{\sigma\left(x_2,x_1\right)
+i\epsilon}dx_2^{0}dx_1^{0}\equiv \eta_{{\bar \mu}{\bar \nu}}\,\, I\left(\mathbf{2},\mathbf{1}\right).
\end{equation}
From  expression (\ref{def1potef}) for ${\cal W}^{\left(\mathbf{2},\mathbf{1}\right)}$ we will obtain the effective
interaction operator, after evaluating the  integrals
in (\ref{def3pe}) and after choosing an ordering
prescription to complete the construction.
To first order in the curvature, the poles of the Synge function are slightly
displaced along the real line with respect to those of the flat expression. Because of this, we can take
in  ({\ref{def3pe}) the same integration contour used in
the Minkoswki spacetime calculation, but inserting the modified poles. The result of the
contour integration is
\begin{eqnarray}\label{Iw}
I\left(\mathbf{2},\mathbf{1}\right)= -i\delta\left(\Omega\right)
\frac{e^{iw\frac{b_0}{\left(1-a_{00}\right)}}}{\left|\rho\right|\left(1-a_{00}\right)}
\left[\theta\left(w\right)e^{iw\left|\rho\right|}+\theta\left(-w\right)e^{-iw\left|\rho\right|}
\right],
\end{eqnarray}
where
\begin{equation}
\label{wOmedefs}
w \equiv \Delta_{fi}^{\left(2\right)}-\Delta_{fi}^{\left(1\right)}, \quad
 \quad \left|\rho\right|
\circeq   r\left[1+\frac{1}{2}\left(a_{00} +\frac{\xi_{h}}{r^2}\right)\right].
\end{equation}
We  have introduced the additional  notation
\begin{eqnarray}
&& \qquad \qquad r^2=\delta_{ab}\left(x_2-x_1\right)^{a}\left(x_2-x_1\right)^{b}, \quad
a_{00}={\cal Q}_{0a0b}{\cal A}^{ab},\\
&& b_{0}={\cal Q}_{0acb}{\cal A}^{ab}\left(x_2-x_1\right)^{c}, \quad
\xi_{h} ={\cal Q}_{cadb}{\cal A}^{ab}\left(x_2-x_1\right)^{c}\left(x_2-x_1\right)^{d}.
\end{eqnarray}
Here ${\cal A}^{ab}$ is obtained from Eq. (\ref{Aabdef}) by making the replacements $x' \rightarrow x_1, \,  x \rightarrow x_2$.
The study of the symmetry properties under the exchange ${\mathbf 1} \leftrightarrow \mathbf{2}$ in  Eqs. (\ref{SfiW}) and (\ref{def1potef}) leads to the property
$I\left(\mathbf{2},\mathbf{1}\right)=I\left(\mathbf{1},\mathbf{2}\right)$, which can be directly verified in  expression (\ref{Iw}).
Next we expand both exponentials up to order  $w^{2}$ obtaining
\begin{eqnarray}\label{2FTGmunu}
&&{\cal G}_{{{\bar \mu}}{{\bar \nu}}}= -i\delta\left(\Omega\right)\eta_{{{\bar \mu}}{{\bar \nu}}}
\left\{\frac{1}{r}\left[ 1+\frac{1}{2}\left(a_{00}-\frac{\xi_{h}}
{r^2}\right)\right] +i\left(\frac{w \,b_{0}}{2r}\right) \right. \nonumber \\
&&\qquad \qquad \left.-\frac{r}{2}\left(\frac{w^{2}}{4}\right)\left(1+\frac{1}{2}\left(3a_{00}+\frac{
\xi_{h}}{r^2}\right)\right)\right\}.
\end{eqnarray}
The above expression still contains terms of order higher than first in
the curvature because
$w \circeq  w_{0}+w_{h}$.  The final reduction to the desired order  will be obtained after we rewrite the powers of $w$ as
functions of the one-particle  Hamiltonian in the presence of a gravitational background
\begin{eqnarray}\label{Hlibre}
&&\qquad  \hat{\cal {H}}\circeq   -\frac{c}{4}\gamma^{\bar{a}}\gamma^{\bar{b}}h_{0a}\hat{p}_{b}+
c\gamma^{\bar{0}}\gamma^{\bar{a}}\left(-\hat{p}_{a}+\frac{1}{2}h^{c}{}_{a}\hat{p}_{c}
+\frac{1}{2}h_{00}\hat{p}_{a}\right)\nonumber  \\
&&-i\hbar c\gamma^{\bar{0}}\gamma^{\bar{\mu}}\Gamma_{\mu}\left(x\right)
-\frac{3c}{4}h^{a}{}_{0}\hat{p}_{a}+imc^{2}\left(1-\frac{1}
{2}h_{00}\right)\gamma^{\bar{0}}
+\frac{i}{4}mc^{2}h_{0a}\gamma^{\bar{a}},
\end{eqnarray}
which is hermitian in the curved inner product
\begin{equation}\label{innerprod}
\langle \phi,\psi\rangle
=-\int \phi^{T \ast}\, \gamma^{\bar{0}}\gamma^{0}\, \psi \, \sqrt{-g}\,\, d^{3}x.
\end{equation}
Let us recall that in Eq. (\ref{Hlibre}) we have expanded all curved quantities up to first order in the curvature, which explains the contraction between
tetrad and coordinate indexes that appears. This feature will frequently repeat itself in what follows.
\section{The effective relativistic interaction}
The effective interaction operator $\hat{{\cal W}}_{e}$ is constructed
in such a way that the $w$-dependent contributions in (\ref{2FTGmunu})  arise  as a consequence of taking the matrix elements
of  $\hat{{\cal W}}_{e}$ indicated in the right hand side of Eq. (\ref{SfiW}). The operator $\hat{{\cal W}}_{e}$ will be a matrix operator acting on the quantum-mechanical Hilbert space of the system.  To perform such construction we  use the identification ${\hat p}_a= -i\hbar \partial_a$ in (\ref{Hlibre}), together with
the eigenvalue equation for each spinor and  Eq. (\ref{2FTGmunu}). We start by substituting
(\ref{2FTGmunu}) in (\ref{SfiW}) in such  a way  that, in the inner product (\ref{innerprod}), we rewrite the
transition element as
\begin{equation}\label{Weffdefinal}
S_{fi}=i\delta(\Omega)
\langle\Phi_{f}\vert\hat{O}\left[\phi_{C}+\left(\frac{iw}{2}\right)\, F_{1}
+\frac{1}{2}\left(\frac{iw}{2}\right)^{2}r\left(1+F_{2}\right)\right]\vert\Phi_{i}\rangle,\nonumber \\
\end{equation}
with the additional definitions
\begin{eqnarray}
&&\label{FsandOdef}
\hat{O}\equiv\eta_{{\bar \mu}{\bar \nu}}\left( \frac{\gamma^{0}_{\left(2\right)} \gamma^{{\bar{\mu}}}_{\left(2\right)}}
{g^{00}\left({\mathbf
x_2}\right)}\right) \left(\frac{
\gamma^{0}_{\left(1\right)}\gamma^{{\bar{\nu}}}_{\left(1\right)}}{g^{00}\left({
\mathbf x_1}\right)}\right) ,\quad
\phi_{C}\equiv\frac{1}{r}\left[1+\frac{1}{2}\left(a_{00}-\frac{\xi_{h}}{r^2}\right)\right], \\
&& \qquad \qquad \qquad \qquad F_1=\frac{b_0}{r}, \quad F_2=\frac{1}{2}\left(3 a_{00}+\frac{\xi_h}{r^2}\right) .
\end{eqnarray}
Most applications will require to consider the low velocity limit of both particles, which arises from the gamma matrices contributions
 to the operator $\hat{O}$ after  taking the non-relativistic approximation. In order to keep track of the desired  order in the
approximation, it is convenient to expand $\hat{O}$ in powers of
\begin{equation}\label{VelOper}
\mathbf{A}_{(n)}^{{\bar k}} \equiv-\gamma_{(n)}^{\bar{0}}\gamma_{(n)}^{\bar{k}},
\end{equation}
which will produce terms of  order ${{v}_{(n)}^{{\bar k}}}/{c}$ in the corresponding limit. We consider this expansion up to second order.
To preserve hermiticity in the inner product (\ref{innerprod}) our construction must keep the expansion of all quantities up to  first order in the curvature. In this way,  the required  expression for $\hat{O}\circeq \sum_{A=0}^2 \, \hat{O}_{A}$ is given by
\begin{eqnarray}
&&\hat{O}_{0}\equiv-{\mathbf I}_{4 \times 4}\left[1-\frac{1}{2}\left(h_{00}^{(2)}+h_{00}^{(1)}\right)\right], \quad
\hat{O}_{-2}\equiv\left[1-\frac{1}{2}\left(h_{00}^{(2)}+h_{00}^{(1)}\right)\right]
\left(\mathbf{A}^{\bar a}_{(2)}\mathbf{A}^{\bar a}_{(1)}\right) \nonumber \\
&&\label{Op1def}
\hat{O}_{-1}\equiv-\frac{1}{4}\left\{\left(h_{0a}^{(2)}+h_{0a}^{(1)}
\right)\left(\mathbf{A}^{{\bar a}}_{(2)}+\mathbf{A}^{{\bar a}}_{(1)}\right) -i\epsilon^{{\bar c}{\bar d}}{}_{{\bar a}}\left[\mathbf{A}^{{\bar a}}_{(2)}\left(h_{0c}{\mathbf \Sigma}_{{\bar d}}\right)_
{(1)}+\mathbf{A}^{{\bar a}}_{(1)}\left(h_{0c}{\mathbf \Sigma}_{{\bar d}}\right)_{(2)}\right] \right\},\nonumber \\
\end{eqnarray}
where the subindex $A=0, -1, -2$ denotes the power  $c^{A}$ to be achieved in  the non-relativistic approximation.\footnote{ We have introduced the notation: $h_{\mu\nu}^{(n)}\equiv h_{\mu\nu}\left(\mathbf{x}_{n}\right),\, n=1,2$ and  $\left[\mathbf{A}_{(n)}^{{\bar a}},\mathbf{A}_{(n)}^{{\bar b}}\right]\equiv-2i\epsilon^{{\bar a} {\bar b} {\bar c}}\mathbf{\Sigma}_{(n){\bar c}}$.} Substituting (\ref{Op1def}) in (\ref{Weffdefinal}), rewriting $w$ as $\omega\left(\hbar c\right)^{-1}$, where $\omega$ is associated to the difference of energy eigenvalues, we find
\begin{equation}\label{appconsCNF}
S_{fi}\approx i \delta(\Omega)\,
\langle\Phi_{f}\vert\left(\hat{W}_{0}+\hat{W}_{-1}+\hat{W}_{-2}\right)|\Phi_{i}\rangle,
\end{equation}
with
\begin{eqnarray}
\label{WC1opdef}
\hat{W}_{0}\equiv\alpha_{e}\, \hat{O}_{0}\, \phi_{C}, \qquad
\hat{W}_{-1}\equiv\alpha_{e}\left[\hat{O}_{-1}\, \phi_{C}+\hat{O}_{0}\left(\frac{i\omega}{2\hbar
c}\right)F_{1}\right], &&\\
\label{W2opdef}
\hat{W}_{-2}\equiv \alpha_{e}\left[\hat{O}_{-2}\, \phi_{C}+\hat{O}_{-1}\left(\frac{i\omega}{2\hbar
c}\right)F_{1}-\frac{I_{4 \times 4} }{2}\left(\frac{i\omega}{2\hbar c}\right)^{2}r\left(1+F_{2}\right)\right].&&
\end{eqnarray}
 We must now choose a prescription in order to convert each term of (\ref{appconsCNF}) into a hermitian operator acting on the quantum-mechanical Hilbert space of the system, in particular upon the wave functions in (\ref{appconsCNF}). Following Greiner \cite{Greiner}, we take
\begin{eqnarray}
\label{omegpresc}
\omega\hat{R}\to\hat{{\cal R}}=
\left[\hat{{\cal H}}^{\left(2\right)}-\hat{{\cal H}}^{\left(1\right)},\hat{R}
\right], \quad
\omega^{2}\hat{U}\to\hat{{\cal U}}=
\left(2i\right)^{2}
\left[\hat{{\cal H}}^{\left(2\right)},\left[\hat{{\cal H}}^{\left(1\right)},\hat{U}\right]\right],
\end{eqnarray}
where $\hat{R}$ and $\hat{U}$ are arbitrary operators,  hermitian in the curved inner product,   and  $\hat{{\cal H}}^{\left(n\right)}$
is the  free Dirac Hamiltonian (\ref{Hlibre}) for the $n$-th particle. The construction (\ref{omegpresc}) relies on the fact that both $|\Phi_i\rangle$, $|\Phi_f\rangle$ are eigenstates of the Hamiltonian (\ref{Hlibre}) for each particle.
Before calculating  the corresponding commutators we introduce some compact notation
\begin{eqnarray}
&&\partial_{a}^{\left(1,2\right)}F_ {2} \equiv\frac{1}{r}\Theta_{a}^{(1,2)},\quad
\partial_{a}^{\left(2\right)}\partial_{b}^{\left(1\right)}F_{2}\equiv\frac{1}{r^{2}}{\cal B}_{ab}, \quad \partial_a^{(n)}=\frac{\partial}{\partial x^a_n},\\
&&\mathbf{r}\equiv\mathbf{x}_2-\mathbf{x}_{1},\quad r=|\mathbf{r}|, \quad \hat{\mathbf{n}}=\frac{\mathbf{r}}{r},\quad
\tilde{a}_{00} \equiv a_{00}-h_{00}^{(2)}-h_{00}^{(1)},\\
&& \qquad \quad z^{a}\equiv x^{a}_{2}+x^{a}_{1},\quad  \tilde{F}_{2} \equiv F_{2}-\frac{1}{2}\left(h_{00}^{(1)}+h_{00}^{(2)}\right).
\end{eqnarray}
Following (\ref{omegpresc}), we construct the operators $ \hat{{\cal R}}: \hat{{\cal W}}_{e}^{C},\, \hat{{\cal W}}_{e}^{CU}$ and $\hat{{\cal W}}_{e}^{B}$ arising  from
$\hat{R}:  \hat{W}_{0}, \, \hat{W}_{-1}$ and $\hat{W}_{-2}$, respectively. The final effective two-body interaction due to  one-photon exchange, which  includes retardation effects up to second order,  is
\begin{equation}\label{effintOp}
\hat{{\cal W}}_{e}\circeq  \hat{{\cal W}}_e^{C}+\hat{{\cal W}}_{e}^{CU}+\hat{{\cal W}}_{e}^{B}.
\end{equation}
By construction $\hat{{\cal W}}_{e}$  is hermitian in the scalar product (\ref{innerprod}) and we
identify it with the interaction energy of the two charges, as seen by the inertial observer.
After a long and tedious calculation,  keeping  only linear  terms in the curvature, we find
\footnote{The notation is: $  ({\mathbf h} \times {\mathbf \Sigma})^a
 =\epsilon^{{\bar a}{\bar b}  {{\bar c}}} h_{0b}{\mathbf \Sigma}_{\bar c}$. Since the results are to first order in the curvature, indices of quantities which are factors of $h_{\mu\nu}$
are lowered or raised by the flat-space metric.}
\begin{eqnarray}
\label{WeFormaFinalFinal}
&&\qquad \qquad \qquad \quad  \frac{\hat{\mathcal{W}}_{e}}{(-\alpha _{e})}=\frac{1}{2r}\left[ 1+\frac{1%
}{2}\left( \tilde{a}_{00}-\frac{\xi _{h}}{r^{2}}\right) \right] \;\mathbf{I}%
_{4\times 4}  \nonumber \\
&&-\frac{1}{2}\mathcal{Q}_{0abd}\left[ \frac{\mathcal{A}^{ab}}{r}\left(
\delta ^{cd}-\;\hat{n}^{c}\hat{n}^{d}\right) -\frac{r}{12}\;\hat{n}^{a}\hat{n%
}^{b}\delta ^{cd}\right] \mathbf{A}_{(2)}^{{\bar{c}}}
+\frac{1}{8}\mathcal{Q}_{0acb}\left[ {\mathbf A}_{(2)}^{\bar{a}}z^{b}+z^{a}{\mathbf A}_{(2)}^{%
\bar{b}}\right] \hat{n}^{c} \nonumber \\
&& +\frac{1}{4r}\mathbf{A}_{(2)}^{{\bar a}}\left( \;{%
h}^{(1)}_{0b}\left[ 2\hat{n}^{a}\hat{n}^{b}-3\delta ^{ab}\right] -{h}^{(2)}%
_{0b}\delta ^{ab}\right) +\frac{1}{8r}\mathbf{A}_{(2)}^{{\bar a}}\left[ \hat{n}%
^{a}\hat{n}^{b}-3\delta ^{ab}\right] \left( \mathbf{h}\times i\mathbf{\Sigma
}\right) _{(1)}^{b}\nonumber  \\
&&\qquad \qquad -\frac{1}{4r}\mathbf{A}_{(2)}^{{\bar{b}}}\left[ \left( 1+\tilde{a}_{00}-%
\frac{\xi _{h}}{r^{2}}-\tilde{F}_{2}\right) \delta _{bc}+\left( 1+\tilde{F}%
_{2}\right) \hat{n}_{b}\hat{n}_{c} \right. \nonumber \\
&&\qquad \qquad \left.+ \left(\delta^{bd}+2\hat{n}^{b}\hat{n}^{d}\right){h}^{(1)}_{dc}
+2\hat{n}^{b}{{\Theta }}^{(1)}_{c}+%
\mathcal{B}_{bc}\right] \mathbf{A}_{(1)}^{{\bar{c}}}+\left( {\mathbf 1} \leftrightarrow
{\mathbf 2} \right)
\end{eqnarray}
 Equation (\ref{WeFormaFinalFinal}) constitutes the main result of this work. In the case of zero gravitational interaction, (\ref{WeFormaFinalFinal}) reduces to the flat space Coulomb plus Breit interaction. Contrary to the flat spacetime case, $\hat{\mathcal{W}}_{e}$ contains terms of order $c^{-1}$ arising from the contributions proportional to ${\mathbf A}_{(n)}^{\bar a}$. Those terms  are  essential in the hermiticity of the complete interacting term. The corrected  effective potential exhibits two special features, which were absent in previous discussions of the problem: (i) the coupling of spatial and spin degrees of freedom and (ii) the coupling of the standard center of mass and relative coordinates. These effects, which  also appear  in the case of two charged particles in the presence of an external inhomogeneous magnetic field \cite{Lesanovsky-Schmiedmayer-Schmelcher}, may offer additional  possibilities for gravitational testable effects at the quantum level.
 Using (\ref{WeFormaFinalFinal}) we  construct the relativistic Hamiltonian for two charges in a background gravitational field, including gravitational corrections due to one-photon exchange, as
\begin{equation}\label{H12final}
\hat{\mathfrak{H}}\left(\mathbf{1},\mathbf{2}\right)=\hat{{\cal H}}_{(1)}+\hat{{\cal H}}_{(2)}+\hat{{\cal W}}_{e},
\end{equation}
where Eq. (\ref{Hlibre}) gives each free Dirac Hamiltonian $\hat{{\cal H}}$.

\section{Final Comments}
Summarizing, for an observer attached to a  freely falling frame outside the gravitational sources, we have constructed a quantum-mechanical two-body relativistic Hamiltonian describing the electromagnetic interaction between two charges in the presence of an arbitrary  weak background gravitational field. Such Hamiltonian  takes into account retardation effects up to  order $(v/c)^2$ and it is hermitian in the curved scalar product (\ref{innerprod}). The gravitational corrections are calculated only up to terms linear in the curvature.
In order to extract physical consequences of the above Hamiltonian it is usually necessary to take the non-relativistic approximation. In our case this will require to perform a two-body Foldy-Wouthuyzen transformation \cite{Chraplyvy}. Even though this transformation can be performed in the full Hamiltonian (\ref{H12final}), the result is rather involved and does not shed much light unless one considers  a particular situation of physical interest \cite{TesisAlex}. When dealing with applications, our general strategy will be  to start by selecting  and  estimating  the order of magnitude of  the many competing terms that appear in (\ref{H12final}). This will allow us to isolate the dominant contributions of the gravitationally induced terms, which will be subsequently considered as perturbations of the resulting zeroth-order Hamiltonian.
In many  situations our observer would be  interested in a gas of  atoms
at temperature $T$ located at some distance $l_{R}$ from the origin of the
freely-falling frame.  Five lengths are relevant and must be estimated: the local curvature radius $D$,
the typical size of the
coordinate patch $l_{O}$, the estimated position of the atom $l_{R}$, the inter-particle distance among the charges $a_{0}$,
and the de Broglie wavelength of the atom $\lambda_{v}$. The dimensionless parameters that measure the gravitational effects are
$
\bar{\gamma}^{2}=\left({l_{O}}/{D}\right)^{2},\, \gamma^{2}=\left({l_{R}}/{D}\right)^{2},\,
\gamma_{0}^{2}=\left({a_{0}}/{D}\right)^{2},\, \breve{\gamma}^{2}=\left({\lambda_{v}/D}
\right)^{ 2},\,
$
where we take $a_{0}$ to be the  Bohr radius of the atom. We certainly have $ \gamma^2, \gamma_0^2, {\breve \gamma}^2 <<< {\bar \gamma}^2 << 1$. Nevertheless, in comparison with the one-particle formulation of the problem, the presence of the additional parameter $ \gamma^2$, which is related to the location of the atom, allows us to probe regions where the tidal forces become relevant. Even  considering} its expected small value, this parameter could provide some amplifying  factor leading to situations where  observability might be enhanced. Given that we require to perform a non-relativistic approximation in our two-body relativistic Hamiltonian (\ref{H12final}), we must also consider additional dimensionless parameters related to the  velocity of the atom $\eta \equiv V_{R}\cdot c^{-1}$ and to the velocity of the electron $\eta_{0}\sim\alpha_{e}\approx7.3\times10^{-3}$. Under solar system conditions, for $T \sim {10^{4}-10^{2}}\, {\rm K}$, we find $\eta \sim 10^{-5}-10^{-6}$. Using these  values of the parameters we can make  further estimations: for astronomical observations  within the solar system $\gamma^2 \geq \eta\eta_0$, for strong gravitational fields near the origin $\gamma_0^2$ might become the dominant parameter, and  for very low temperature regimes ${\breve \gamma}^2$ could become relevant.
Specific applications are beyond the scope of this letter and will be considered in  forthcoming publications \cite{TesisAlex,ACLU2}.

\section*{Acknowledgements}

JAC and LFU would like to thank useful discussions with C. Chryssomalakos   and  D. Sudarsky. E. Nahmad is gratefully acknowledged for a careful reading
of the manuscript.  LFU
is partially supported by projects CONACYT \# 55310 and DGAPA-UNAM-IN111210. He also
acknowledges support from RED-FAE, CONACYT. JAC has been partially supported by projects CONACYT \# 55310,
DGAPA-UNAM-IN109107 and a DGEP-UNAM scholarship.

\begin{figure}[b!]
\begin{center}
\includegraphics[scale=0.3]{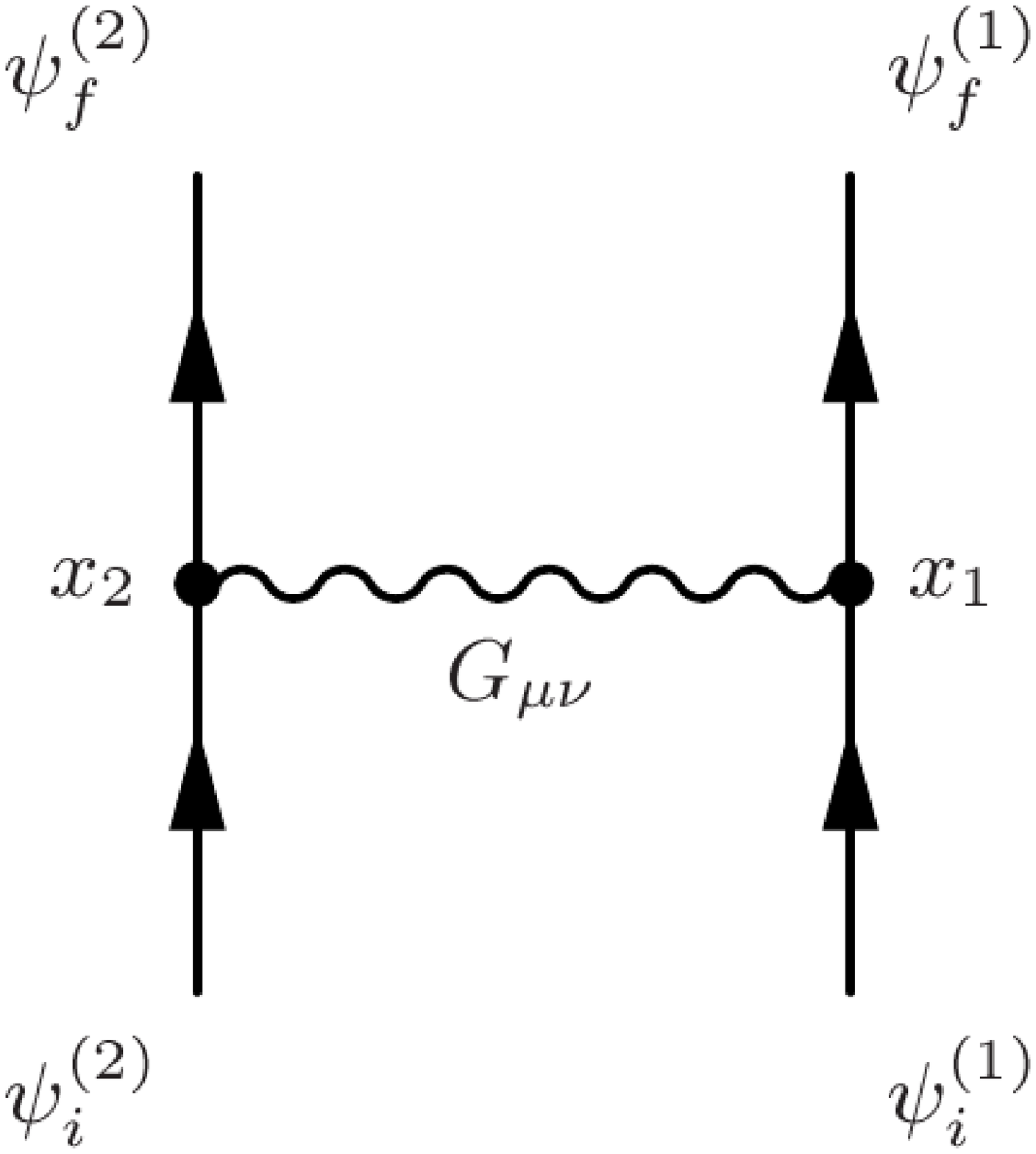}
\caption{The one-photon exchange diagram leading to the effective
interaction potential. Fermions and photons include the effects of the
gravitational background.}\label{fig:proceso}
\end{center}
\end{figure}

\end{document}